\documentclass[twocolumn,trackchanges]{aastex7}
\usepackage{subcaption} 
\usepackage{tabularx}
\usepackage{amsmath}
\usepackage{xfrac}

\newcommand{\erosita}{\textit{eROSITA}}
\newcommand{\chandra}{\textit{Chandra}}
\newcommand{\xmm}{\textit{XMM-Newton}}
\newcommand{\half}{\sfrac{1}{2}}
\newcommand{\sptplanck}{{\it Planck}/SPT}
\newcommand{\planck}{{\it Planck}}
\newcommand{\xrism}{{\it XRISM}}

\begin{document}

\title{Intracluster Medium Fluctuations on Scales up to 1 Mpc: A Combined eROSITA and SPT/Planck Analysis of Abell 3266}

\author[orcid=0009-0006-8186-7440]{H. Saxena}
\affiliation{California Institute of Technology, 1200 East California Boulevard, Pasadena, California, USA}
\email[show]{hsaxena@caltech.edu}  

\author[orcid=0000-0002-7726-4202]{A. Heinrich} 
\affiliation{Department of Astronomy \& Astrophysics, University of Chicago, 5640 S Ellis Avenue, Chicago, IL 60637, USA}
\email{amheinrich@uchicago.edu}

\author[orcid=0000-0002-8213-3784]{J. Sayers}
\affiliation{California Institute of Technology, 1200 East California Boulevard, Pasadena, California, USA}
\email{jsayers@astro.caltech.edu}

\author[orcid=0000-0001-7630-8085]{I. Zhuravleva}
\affiliation{Department of Astronomy \& Astrophysics, University of Chicago, 5640 S Ellis Avenue, Chicago, IL 60637, USA}
\email{zhuravleva@uchicago.edu}

\author[orcid=0000-0002-7619-5399]{E. Bulbul}
\affiliation{Max-Planck-Institut fur extraterrestrische Physik (MPE), Gießenbachstraße 1, D-85748 Garching bei Munchen, Germany}
\email{ebulbul@mpe.mpg.de}

\author[orcid=0000-0003-2189-4501]{J. Sanders}
\affiliation{Max-Planck-Institut fur extraterrestrische Physik (MPE), Gießenbachstraße 1, D-85748 Garching bei Munchen, Germany}
\email{jsanders@mpe.mpg.de}

\author[orcid=0000-0001-8868-0810]{C. Avestruz}
\affiliation{Department of Physics and Leinweber Institute for Theoretical Physics,
University of Michigan, 450 Church St, Ann Arbor, MI 48109}
\email{cavestru@umich.edu}

\author[orcid=0000-0002-3351-3078]{R. Basu Thakur}
\affiliation{NASA Jet Propulsion Laboratory, 4800 Oak Grove Dr., Pasadena, CA 91011, USA}
\email{ritoban@caltech.edu}

\author[orcid=0000-0001-5210-7625]{E. Battistelli}
\affiliation{Physics Department, Sapienza University of Rome, Piazzale Aldo Moro 5, 00185 Rome, Italy}
\email{Elia.Battistelli@roma1.infn.it}

\author[orcid=0000-0002-9325-1567]{A. Botteon}
\affiliation{INAF – IRA, via P. Gobetti 101, 40129 Bologna, Italy}
\email{botteon@strw.leidenuniv.nl}

\author[orcid=0009-0007-6324-479X]{F. Cacciotti}
\affiliation{Physics Department, Sapienza University of Rome, Piazzale Aldo Moro 5, 00185 Rome, Italy}
\email{federico.cacciotti@roma1.infn.it}

\author[orcid=0000-0002-5127-0401]{F. Columbro}
\affiliation{Physics Department, Sapienza University of Rome, Piazzale Aldo Moro 5, 00185 Rome, Italy}
\email{fabio.columbro@roma1.infn.it}

\author[orcid=0000-0003-4522-5214]{A. Coppolecchia}
\affiliation{Physics Department, Sapienza University of Rome, Piazzale Aldo Moro 5, 00185 Rome, Italy}
\email{alessandro.coppolecchia@roma1.infn.it}

\author[orcid=0000-0001-5493-5475]{S. Cray}
\affiliation{University of Minnesota, 115 Union St. SE, Minneapolis, MN 55455, USA}
\email{scray@umn.edu}

\author[orcid=0000-0001-6547-6446]{P. de Bernardis}
\affiliation{Physics Department, Sapienza University of Rome, Piazzale Aldo Moro 5, 00185 Rome, Italy}
\email{paolo.debernardis@roma1.infn.it}

\author[orcid=0000-0001-7859-2139]{M. De Petris}
\affiliation{Physics Department, Sapienza University of Rome, Piazzale Aldo Moro 5, 00185 Rome, Italy}
\email{marco.depetris@roma1.infn.it}

\author[orcid=0000-0002-3714-1507]{L. Lamagna}
\affiliation{Physics Department, Sapienza University of Rome, Piazzale Aldo Moro 5, 00185 Rome, Italy}
\email{luca.lamagna@roma1.infn.it}

\author[orcid=0000-0001-8914-8885]{E.T. Lau}
\affiliation{Department of Physics, Nara Women's University, Kitauoyanishimachi, Nara City, Nara 630-8506, Japan}
\email{ethlau@gmail.com}

\author[orcid=0000-0001-5105-1439]{S. Masi}
\affiliation{Physics Department, Sapienza University of Rome, Piazzale Aldo Moro 5, 00185 Rome, Italy}
\email{silvia.masi@roma1.infn.it}

\author[orcid=0000-0002-8388-3480]{A. Paiella}
\affiliation{Physics Department, Sapienza University of Rome, Piazzale Aldo Moro 5, 00185 Rome, Italy}
\email{alessandro.paiella@roma1.infn.it}

\author[orcid=0000-0002-5444-9327]{F. Piacentini}
\affiliation{Physics Department, Sapienza University of Rome, Piazzale Aldo Moro 5, 00185 Rome, Italy}
\email{francesco.piacentini@roma1.infn.it}

\author[orcid=0000-0003-1560-8580]{E. Rapaport}
\affiliation{California Institute of Technology, 1200 East California Boulevard, Pasadena, California, USA}
\email{erapapor@caltech.edu}

\author[orcid=0000-0001-5636-7213]{L. Rudnick}
\affiliation{University of Minnesota, 115 Union St. SE, Minneapolis, MN 55455, USA}
\email{larry@umn.edu}

\author[orcid=0000-0002-7707-9437]{D. White}
\affiliation{California Institute of Technology, 1200 East California Boulevard, Pasadena, California, USA}
\email{dwhite2@caltech.edu}

\author[orcid=0000-0003-3175-2347]{J. ZuHone}
\affiliation{Center for Astrophysics, Harvard and Smithsonian, 60 Garden St., Cambridge, MA 02138, USA}
\email{john.zuhone@cfa.harvard.edu}

\begin{abstract}
Galaxy clusters form through hierarchical assembly, where smaller substructures merge to build the largest gravitationally bound objects in the universe. These mergers, combined with feedback from AGN, filamentary accretion, and other energy injection processes, generate turbulence and perturbations within the intra-cluster medium (ICM). X-ray and Sunyaev–Zel’dovich (SZ) observations can be utilized to measure these ICM density and pressure inhomogeneities, in turn providing constraints on the effective Equation of State (EOS) of the perturbatons and ICM velocities. In this work, we analyze deep SRG-eROSITA (hereafter \erosita\,) and \sptplanck\ observations of Abell 3266 (A3266), a dynamically complex merging cluster with elongated morphology and significant substructure. We measure pressure and density fluctuations, and compute the power spectra and deprojected 3D amplitudes of these perturbations. We estimate the ratio of pressure-to-density fluctuation amplitudes as $1.00 \pm 0.55$ and non-thermal pressure support $0.068 \pm 0.050$. Density fluctuations are found to be stronger in the northern sector of the cluster compared to the south, consistent with ongoing accretion along a filamentary structure revealed by \erosita. Further, we find the amplitude of density fluctuations increases with radius, qualitatively consistent with the trend found in cosmological simulations. Uncertainties in our results are dominated by the relatively low sensitivity of current \sptplanck\ data, suggesting that improvements in SZ data quality could substantially improve our understanding of ICM energy injection, transport, and dissipation from this technique.
\end{abstract}

\keywords{\uat{High Energy astrophysics}{739}, \uat{Galaxy clusters}{584}, \uat{Intracluster medium}{858}}


\section{Introduction}\label{intro}
\begin{figure*}
    \centering
    \includegraphics[width=\linewidth]{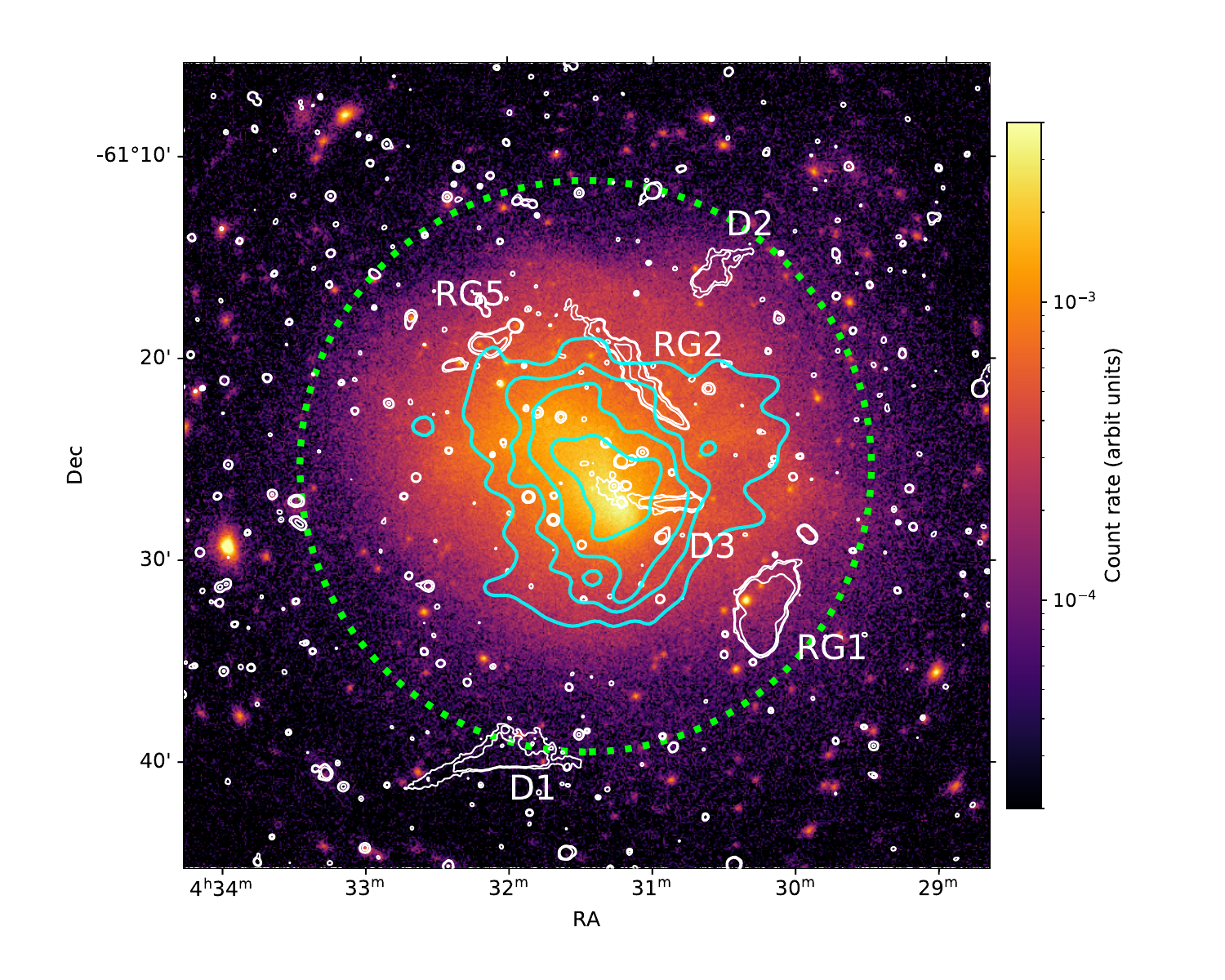}
    \caption{A3266: X-ray image from \erosita\  with SZ contours (cyan) from \sptplanck\ and radio contours (white) from ASKAP \citep{Riseley_2022} overlaid. Distinct radio features are highlighted in white. The dotted circle in green denotes 1 Mpc, the region utilized for the fluctuation analysis in this paper. The X-ray image is lightly smoothed for visualization purposes.}
    \label{fig:sz_xray_radio}
\end{figure*}

Available evidence strongly supports hierarchical structure formation, with larger objects assembling through the merger of smaller substructures \citep{Press_Schechter}. As the most massive collapsed objects in the universe, clusters serve as ideal laboratories to study merger and accretion processes. The intracluster medium (ICM), which contains the majority of a cluster’s baryons, can be probed in X-rays via thermal bremsstrahlung emission \citep{Xray_clusters} and at millimetre wavelengths through the Sunyaev–Zel’dovich (SZ) effect \citep{SZ_effect}. To good approximation from typical observations with existing instruments, these ICM observables trace the line-of-sight integral of the electron density squared and electron pressure, respectively. In the absence of mergers and other processes that inject significant energy into the ICM, and given sufficient time to relax, the ICM asymptotes to smooth density and pressure profiles, consistent with the universality of dark matter halos \citep{Zhuravleva_2012, McDonald_2019}.
\\\\
Accretion shocks associated with the assembly process heat the ICM to  approximately the virial temperature of the galaxy cluster \citep{Ryu2003}. In the absence of other intrinsic energy injection mechanisms, the ICM would radiatively cool, leading to molecular cloud formation and large star formation rates in cluster cores \citep{Fabian_1994}, neither of which is observed in the expected amount from this scenario \citep{Rafferty_2008}. 
There are several sources suggested for the energy injection necessary to offset this radiative cooling, such as jets from the central supermassive black hole \citep{Burns_1990,Churazov_2000,Fabian_2000,McNamara_2007, Perseus_BH} and associated cosmic rays \citep{Sharma_2010, Ruszkowski_2017, Ehlert_2018}, merger driven turbulence and sloshing \citep{Motl_2004,Rossetti_2010, Su_2016, Su_2017, Heinrich_mergers}, and thermal conduction \citep{Narayan_2001,Karen_Yang_2016, Chen_2019}. 
\\\\
The various processes noted above can drive gas motions in the ICM at different locations within the cluster \citep{Dupourqu_2024, Heinrich_mergers, Vazza_2012,ZuHone_2016, Mohapatra_2019, groth2025turbulencesimulatedlocalcluster}. 
Cool-cores of relaxed clusters are often affected by AGN feedback \citep{McNamara_2007, Sotira_2025}, while accretion from the cosmic web and hierarchical mergers dominates gas dynamics elsewhere \citep{Ryu2003,Nelson_2012, lebeau2025velocityfieldsturbulencecosmic}. 
These motions in turn produce fluctuations in ICM thermodynamic properties such as density and pressure. 
The relationship between fluctuations and the underlying velocities is expected to be close to linear \citep{Gaspari_2013,zhuravleva2014}.
Calibrations from cosmological simulations support this conclusion \citep{Irina_Mach,simonte2022}, and thus fluctuations can be utilized to probe the underlying velocity structure and also to assess the overall dynamical state of the cluster \citep{Schuecker_2004,churazov_coma,Gaspari_2013, zhuravleva2014, Mohapatra_2019, Irina_Mach}.
While ICM motions can be measured directly by \xrism{} via emission line shifts and broadening \citep[e.g.,][]{xrism_a2029,xrism_perseus,XRISM_coma} and via the kinetic SZ effect \citep{Sayers2013,Adam_2017}, given current instrumentation, measuring gas motions via density and pressure fluctuations has the added benefit of producing deprojected, multi-scale velocity measurements.
\\\\  
Density fluctuations have been measured using \chandra{} and \xmm{} within $R_\mathrm{2500}$ \citep{Eckert2017,Dupourqu_2024, Heinrich_mergers}.
The large effective area, stable background, and wide field-of-view of \erosita{} make it a promising instrument to measure density fluctuations to approximately twice this radius, i.e., within $R_{500}$, in lower redshift clusters and across a wider range of scales than \xmm{}.
Recently, SZ observations, which are relatively brighter in the outer cluster regions but generally lack the angular resolution to probe small scales, have also been utilized to probe pressure fluctuations, although with marginal detections to date owing to the relatively low signal-to-noise ratio of available data \citep{Khatri_2016,romero_2023,romero2024}.
Complimentary to the X-arithmetic method \citep{Churazov_2016, mccall2025decodingagnfeedbackxarithmetic}, the ratio of pressure to density fluctuations probes the effective equation of state of ICM perturbations \citep{Zhuravleva_2016, Arvalo_2016,Zhuravleva_2018,romero2024}.
This is parameterized as $\zeta=\frac{\delta P/P_k}{\delta\rho/\rho_k}$.
$\zeta$ is expected to be 0, 1 and 5/3 if perturbations appear as isobaric, isothermal, or adiabatic, respectively.
Additionally, pressure fluctuations provide an independent method to constrain multi-scale gas velocities.
\\\\
In this paper, we perform a combined SZ and X-ray analysis of the ICM fluctuations within A3266 using observations from \sptplanck{} and \erosita{}, shown in Fig \ref{fig:sz_xray_radio}, including the first measurement of X-ray fluctuations with the latter instrument. 
A3266 is the most massive member of the Horologium-Reticulum supercluster, with a redshift of 0.0594 and mass $M_{500} = 8.8 \times 10^{14}~M_\odot$ within $R_{500} = 1.43$ Mpc \citep{Ettori_2019_XCOP}. 
It is one of the most X-ray bright galaxy clusters in the sky \citep{Edge_1990, Abell3266_flux}, and as an \erosita{} performance-verification target, A3266 has some of the deepest observations from that instrument \citep{Sanders_eROSITA}.
Previous studies have revealed evidence of a complex merger event, including an elongated and asymmetric surface brightness distribution, and significant substructure within the cluster \citep{XMM_Abell3266_Sauva}. 
\chandra\ observations of the cluster found a cooler filamentary region centered on the central galaxy aligned along a merger axis running northeast \citep{Chandra_Abell3266}. \xmm\ observations suggest that this stripped gas with high metallicity and low entropy is the result of a 1:10 minor merger in a direction close to the plane of the sky \citep{XMM_Abell3266_Fino}.  
A detailed optical spectroscopic structural analysis found that A3266 can be decomposed into a set of six groups and filaments to the north of the cluster, in addition to a cluster core which can be split into two components \citep{Optical_Abell3266}.  
They also conclude that A3266 is not a simple NE-SW merger, but it instead has a range of continuous dynamical interactions taking place, with high velocity dispersion in the core of $v_{\textrm{disp}} \approx 1400$ \citep{Optical_Abell3266}. 
An analysis of deep radio observations by \citet{Riseley_2022}, overlaid in Fig.\ref{fig:sz_xray_radio}, shows the presence of an atypical ‘wrong-way’ relic with strong spectral steepening (D1), a fossil plasma source with a highly curved and ultra-steep spectrum (D2), and a central diffuse ridge (D3) which is the brightest part of a large-scale radio halo \citep{A3266_radiohalo}. 
Further, there are a number of bent-tail radio galaxies, including the wide-angle tail radio galaxy (RG1), a complex radio galaxy with tethers and ribs (RG2, \cite{rudnick2021sourcesourcesribstethers}), and an inclined tail galaxy (RG5). 
Thus, in addition to being a local cluster with accessible SZ and X-ray data, the rich multi-wavelength probes of the merger history of A3266 make it an ideal candidate for such a fluctuation analysis. 
\\\\
This paper is organized as follows. 
Section \ref{data} describes the X-ray and SZ analyses of A3266 using \erosita\ and \sptplanck\ data, including the datasets employed, the $\beta$ model fitting procedure, and the power-spectrum estimation. 
In Section \ref{result}, we combine the X-ray and SZ results to quantify the relative amplitudes of thermodynamic fluctuations, estimate the Mach number and associated non-thermal pressure support arising from these fluctuations, compare fluctuation amplitudes between the northern and southern hemispheres, and present the radial profile of density fluctuations. Finally, Section \ref{impact} explores alternative modeling choices and presents potential improvements from new SZ instrumentation.

\section{Data Analysis} \label{data}

\subsection{SZ data reduction and model fitting} 
\label{sec:sz_analysis}
\begin{figure*}
    \centering
    \begin{subfigure}[t]{0.45\linewidth}
        \centering
        \includegraphics[width=0.97\linewidth]{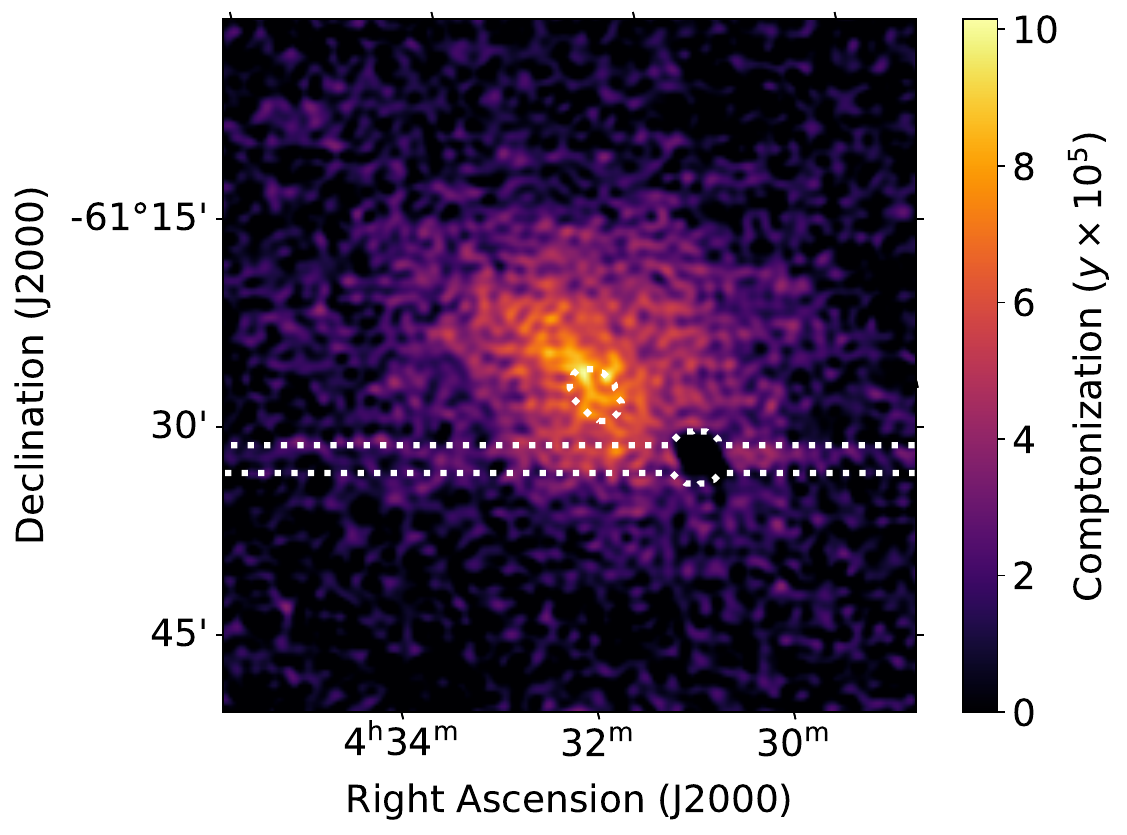}
        \caption{\sptplanck\ $y$-map.} 
        \label{fig:SPT_y_map_masked}
    \end{subfigure}
    \hfill
    \begin{subfigure}[t]{0.45\linewidth}
        \centering
        \includegraphics[width=\linewidth]{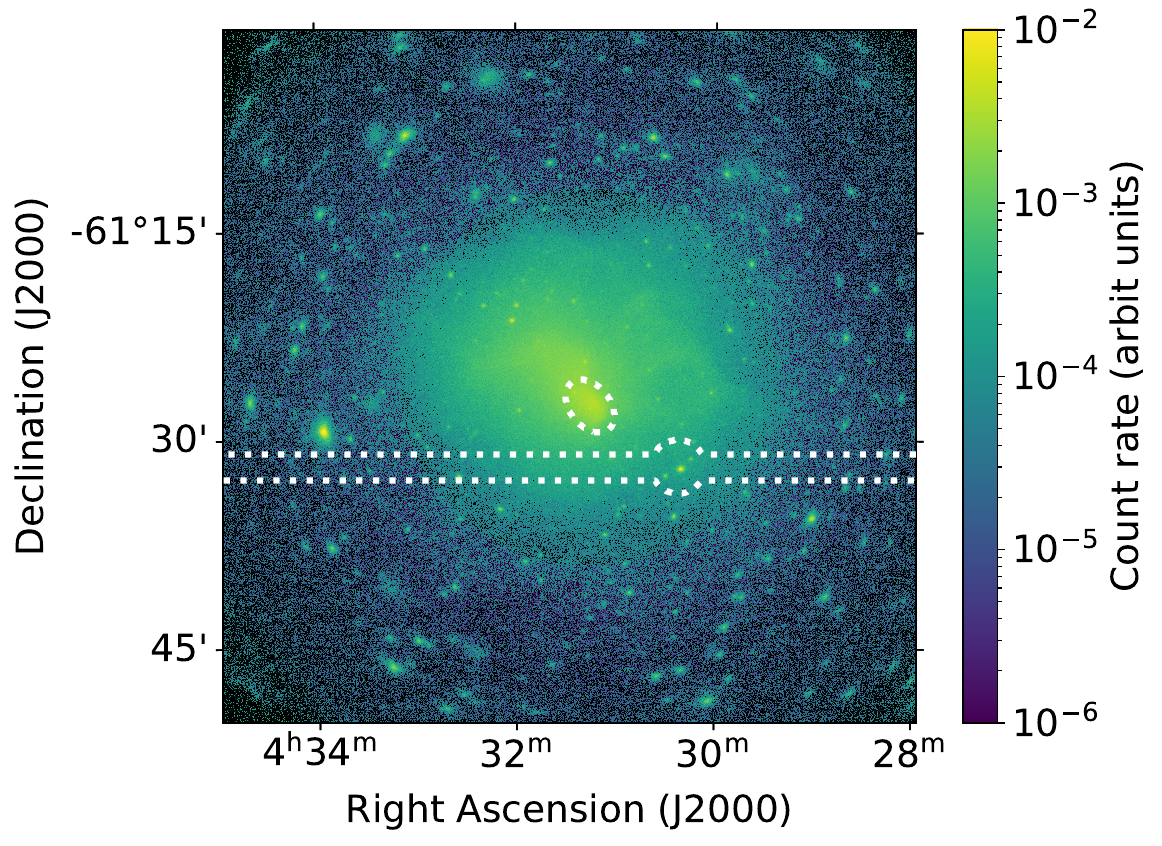}
        \caption{eROSITA X-ray counts map.}
        \label{fig:Xray_countsmap_masked}
    \end{subfigure}

    \caption{
    Images of A3266,
    with masked regions denoted by dotted lines.}
    \label{fig:SPT_Xray_combined}
\end{figure*}

We use the combined South Pole Telescope (SPT) and \planck\ maps released by \citet{Bleem_2015} for our analysis. Since the angular size of A3266 is $2 \times R_{500} = 32\arcmin$, the flat-sky maps in the Sanson-Flamsteed projection are sufficient. 
Following the approach of \citet{romero2024}, who utilized the same combined \sptplanck\ data for a similar fluctuation analysis of different clusters, we perform our analysis with the ``minimum variance'' maps. A $50\arcmin$ diameter cutout of A3266 obtained from this map centered at the X-ray centroid is shown in Fig. \ref{fig:SPT_y_map_masked}. A bright unresolved source is located south-west of the cluster center, identified as SPT-SJ043020-6131.8 in the SPT point source catalog with a SNR of 60.2 \citep{Everett_2020}. We construct a circular mask with radius 2\arcmin\ centered at this bright source ($\text{RA} = 67.585 \ \text{deg}, \ \text{Dec} = -61.530 \ \text{deg}$), and exclude the masked data from our analysis. Further, this source leaves a scan-related artifact at fixed declination spanning the entire range of Right Ascension within our cutout, and so we similarly mask and exclude all data within a strip of total width 2\arcmin. Finally, we also mask and exclude the south-western substructure seen in X-rays, as detailed in Sec.~\ref{sec:x-ray_analysis}. The dotted outline of these masks is shown in Fig. \ref{fig:SPT_Xray_combined}.

\subsubsection{SZ noise estimation}
\label{sec:sz_noise}
To estimate the noise in the SZ data, we obtain identical 50\arcmin\ cutouts from random locations within the \sptplanck\ maps. By repeating all of our analysis steps for an ensemble of these cutouts, we are then able to assess the level of variation in any given parameter value due to noise, fully accounting for any deviations from perfectly un-correlated flat-spectrum map noise. To avoid cutouts with significant astrophysical contamination, we exclude portions of the map that fall behind the masks associated with bright unresolved sources and high dust-emission from the SPT-SZ survey \citep{Bleem_2022}. We further exclude regions within a physical separation of $3 \ \text{Mpc}$ from each object listed in the SPT-SZ cluster catalog \citep{Bocquet_2019}. After applying these exclusions, we obtain 100 cutouts from the map. For any analysis of these cutouts, we also apply the same masks as noted in Sec.~\ref{sec:sz_analysis} for the cutout centered on A3266, to ensure that these random cutouts are treated in an identical manner to the cluster cutout. 

\subsubsection{SZ $\beta$-model fitting}
\begin{figure*}
    \centering
    \begin{subfigure}[t]{0.45\linewidth}
        \centering
        \includegraphics[width=0.96\linewidth]{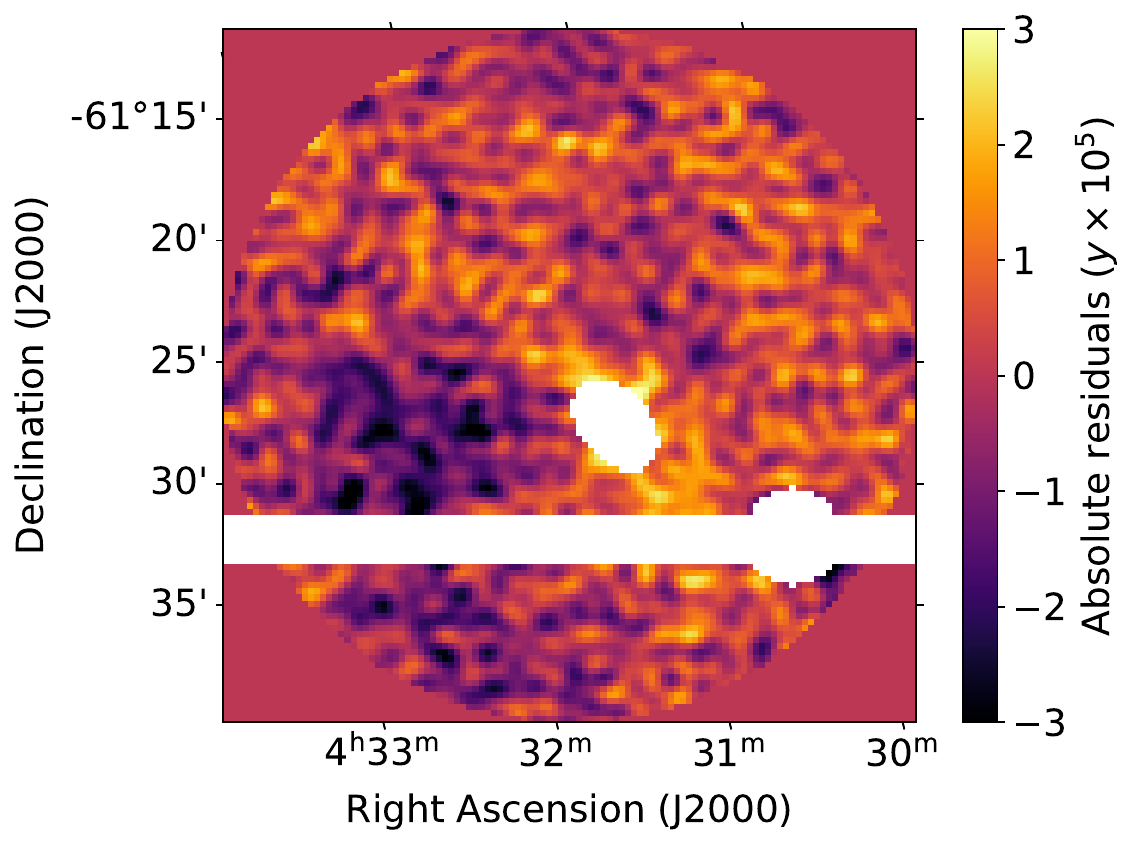}
        \caption{
        \sptplanck\ absolute residuals.}
        \label{fig:abs_SZ_residual}
    \end{subfigure}
    \hfill
    \begin{subfigure}[t]{0.45\linewidth}
        \centering
        \includegraphics[width=\linewidth]{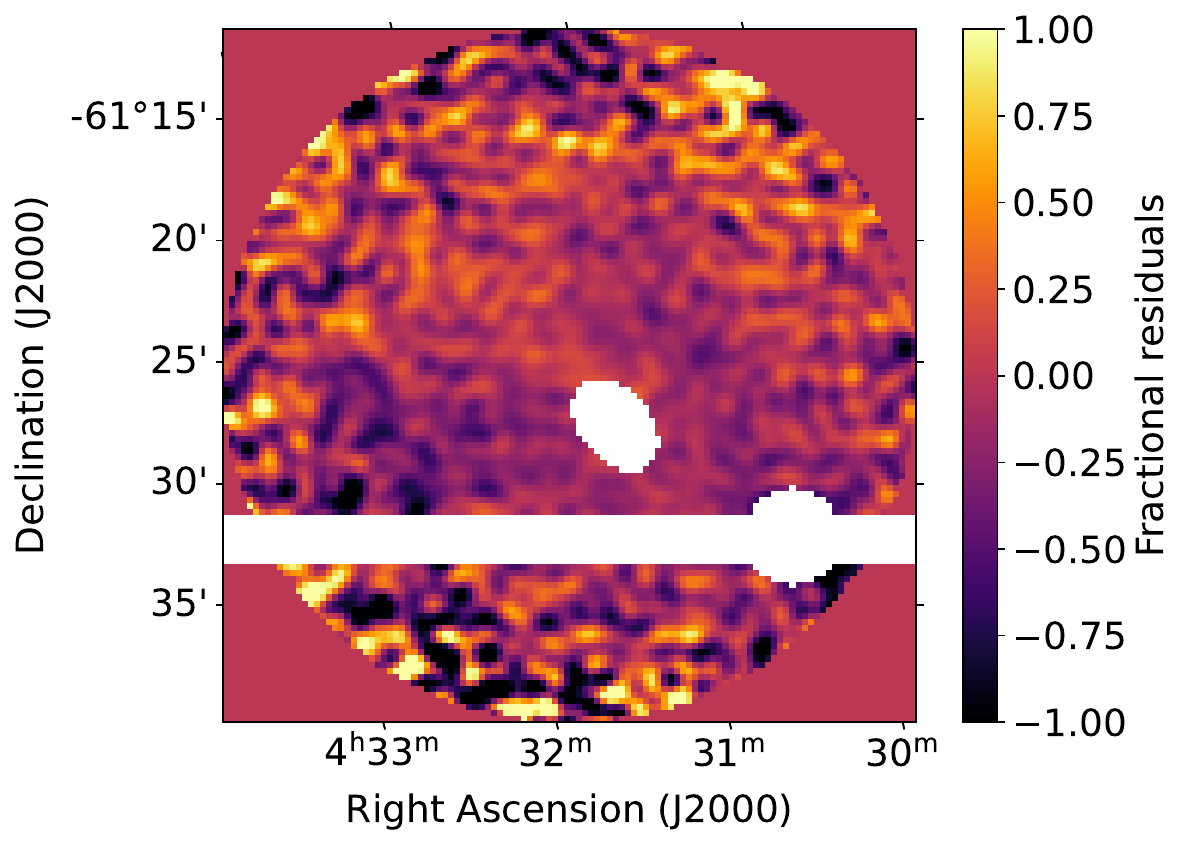}
        \caption{
        \sptplanck\ fractional residuals.}
        \label{fig:frac_SZ_residuals}
    \end{subfigure}

    \vskip\baselineskip 

    \begin{subfigure}[t]{0.45\linewidth}
        \centering
        \includegraphics[width=0.96\linewidth]{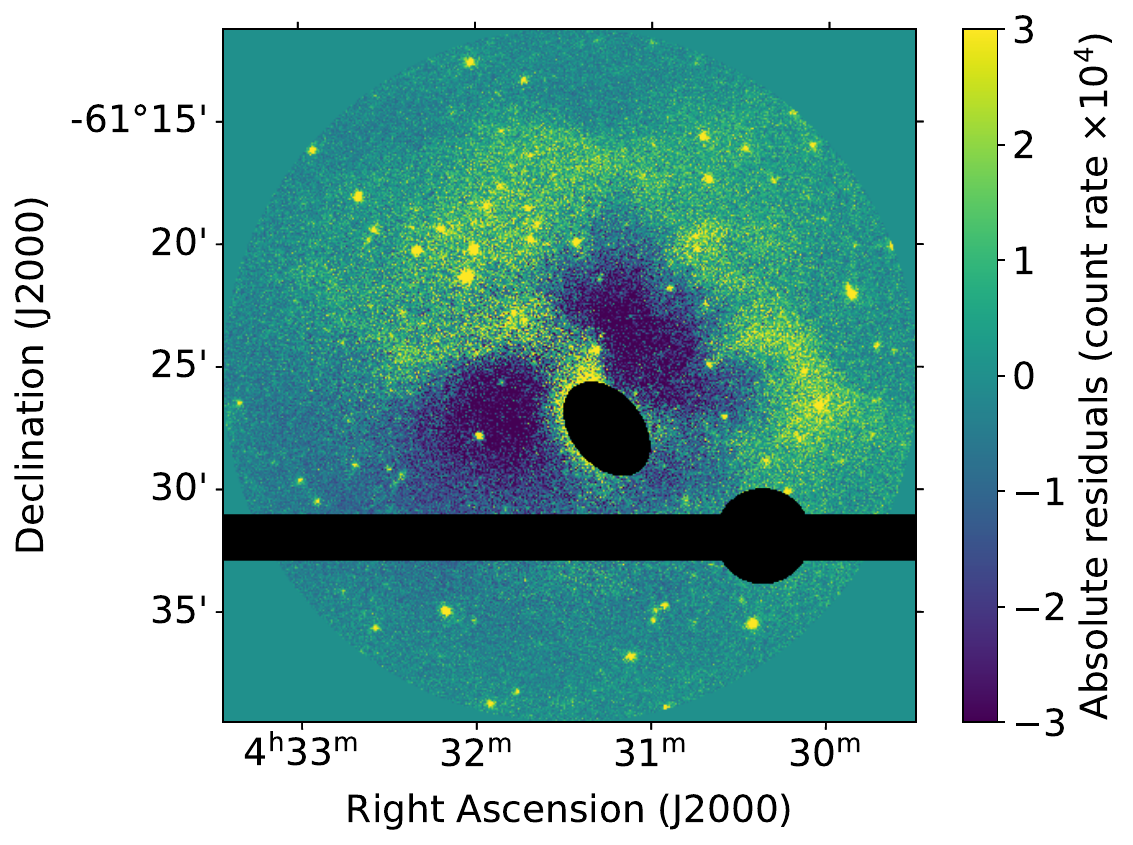}
        \caption{
        \erosita\ absolute residuals.}
        \label{fig:abs_Xray_residual}
    \end{subfigure}
    \hfill
    \begin{subfigure}[t]{0.45\linewidth}
        \centering
        \includegraphics[width=\linewidth]{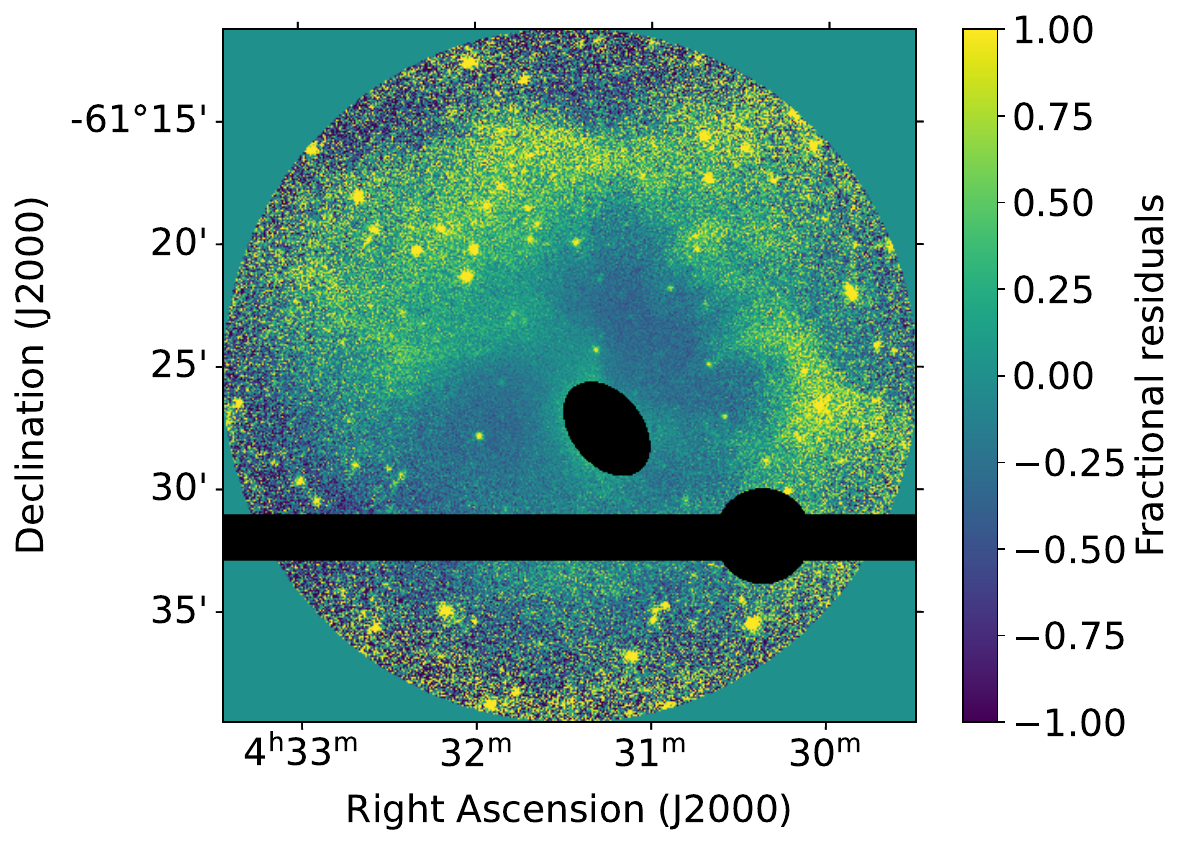}
        \caption{
        \erosita\ fractional residuals.}
        \label{fig:frac_Xray_residual}
    \end{subfigure}

    \caption{
    Residual maps. Absolute residuals are shown after subtracting an elliptical beta-model from the observed data, and illustrate coherent residuals relative to this model in both the SZ and X-ray images, such as the deficit of signal to the SE. Fractional residuals are shown after subtracting, and then dividing by, the patched elliptical $\beta$-model.}
    \label{fig:SPT_Xray_combined_residuals}
\end{figure*}

We fit an elliptical $\beta$-model of the form
\begin{equation}\label{eqn:beta_model}
    y = y_0\ (1 + r_{\text{e}}^2)^{-3\beta + 0.5} + B
\end{equation}
to the 2D map data, where $y_0$ is the overall normalization of the $\beta$-model, $r_{\text{e}}^2 = \bigg( \frac{x_{rot}}{r_c} \bigg)^2 + \bigg( \frac{y_{rot}}{r_c(1-e)} \bigg)^2$ is the normalized elliptical radius from the X-ray centroid where $r_c$ is the core radius, $e$ is the ellipticity, $x_{\text{rot}}$, $y_{\text{rot}}$ are coordinates rotated by an angle $\theta$ relative to the R.A./dec basis, $\beta$ defines the shape of the the model, and $B$ is an overall additive signal offset in the map, which accounts for any potential non-zero local background. The fit is performed using a Markov chain Monte Carlo (MCMC) assuming uniform priors for all the parameters, specifically $[10^{-6}, 10^{-2}]$ for $y_0$, $[0,2]$ for $\beta$, $[0, 25']$ for $r_c$, $[0,1]$ for $e$, $[0, \pi]$ for $\theta$, and $[-1,1]$ for B. We initialize our walkers at starting values of $10^{-5}$ for $S_0$, $10'$ for $r_c$, $0.5$ for $\beta$, $0$ for ellipticity and $\theta$, and $10^{-6}$ for $B$.
We use the \textit{emcee} Python package to run a MCMC over these $4$ dimensions using $50$ walkers over $20,000$ steps, and use the initial half as a burn-in \citep{Foreman_Mackey_2013}. We assume a uniform diagonal map-space noise covariance for these fits, with the diagonal variance values set equal to the mean variance per pixel calculated from the 100 random cutouts described in Sec.~\ref{sec:sz_noise}. We confirm that that the fits are well-converged for all of the parameters, and that the final results are unchanged when the walkers are initialized with different starting values.
\\\\
We then convolve this $\beta$-model with the Gaussian point-spread function (PSF) of the \sptplanck\ maps to generate a symmetric model map for A3266. In accordance with the X-ray analysis described in Sec.~\ref{sec:x-ray_analysis}, and the results from \cite{Heinrich_mergers}, we patch our symmetric model with a fiducial 9\arcmin\ patching scale to remove any large-scale deviation from elliptical symmetry in the cluster. Hereafter, we present our fiducial results from these patched models for both SZ and X-rays. To obtain the fractional residual map, we subtract the model from the map, and then divide the resulting image by the model. Finally, because the signal-to-noise ratio becomes too low to provide meaningful constraints at large radii, we restrict our analysis to $\leq 1$ Mpc ($14.16\arcmin$) from the X-ray centroid, which is accomplished by applying a circular window to our fractional residual image. The absolute (Fig. \ref{fig:abs_SZ_residual}) and fractional (Fig. \ref{fig:frac_SZ_residuals}) residuals from this analysis are shown. 

\subsection{\erosita{} data reduction and model fitting}
\label{sec:x-ray_analysis}
A3266 was observed on 2019-11-11 for 86 ks in the early phase of the \erosita\ observations as a calibration target (Observation ID 700154). In this work, we utilize public data that was processed with software version 001 of the eSASS analysis software \citep{Brunner2022}. The details of the data reduction from the events to the generation of images are described in detail in \cite{Sanders_eROSITA}. After accounting for good time interval filtering (GTI) and removing the flags for corners and intervals with short background flares, the average exposure time per telescope module was 59~ks.
\\\\
Mosaic images, exposure maps, and background images are created using the eSASS tools \texttt{evtool} and \texttt{expmap} using all Telescopes Modules (TMs) in the 0.3-2.3~keV band. For TMs 5 and 7, the softer band is excluded from further analysis to avoid contamination from the light leak \citep{Predehl2021}.
Point sources are identified using the tool \texttt{ermldet} and masked from the image using the procedure described in \cite{erosita_pointsources} and \cite{Sanders_eROSITA}. 
The resultant point sources were masked from the \erosita{} image.
\\\\
Additionally, we mask the bright substructure associated with the infalling subcluster core from the image, and exclude the masked region from our analysis. If this substructure were included in the fluctuation analysis, its sharp surface brightness gradient would dominate the signal, providing little information about fluctuations in the bulk ICM \citep{Heinrich_mergers}. As this substructure is not clearly identifiable in the SZ, it is doubly important to remove in order to properly compare density and pressure fluctuations. To ensure that this subcluster has been fully removed, we iteratively increase the size of the mask, calculating the power spectrum of surface brightness fluctuations for each size (Section \ref{PS}). We find an ellipse approximately 2\arcmin\ by 1.5\arcmin\ is sufficient to remove the substructure's effect from the power spectrum.
\\\\
To construct the unperturbed cluster model, we first determine the ellipticity of the cluster by fitting the exposure-corrected, background-subtracted surface brightness image with a 2D elliptical beta-model of the form described in Equation \ref{eqn:beta_model}. This method struggles to accurately model the entire cluster, resulting in negative residuals in the center of the cluster. It can, however, measure the ellipticity of the cluster. We thus use the ellipticity derived from the 2D fit to measure a radial surface brightness profile of the cluster in elliptical coordinates, which is fit to a single beta-model via a maximum-likelihood fit.
\\\\
The resultant residual image is asymmetric, as the cluster is significantly brighter in the west, likely due to the ongoing merger. The unperturbed model should reflect this large-scale structure. To do so, the elliptical beta-model is patched to correct the asymmetry, which is accomplished by multiplying it by the ratio of the residual image, smoothed with a Gaussian, over the smoothed mask \citep{Zhuravleva_2015}. The details of this method are described in \citet{Heinrich_mergers}, Appendix A. A fiducial patching scale of 9\arcmin\ is sufficient to visually remove the asymmetry; however, as the appropriate patching scale is uncertain, different model choices are explored in more detail in Appendix \ref{sec:model_impact_ps} and their impact is quantified in Section \ref{impact}. 

\subsection{Power Spectra Measurements} \label{PS}
To quantify surface brightness fluctuations, we compute the power spectra of the fractional residual images for both the X-ray and SZ data. Since both maps contain masked regions (e.g., the bright point-like source and substructures within the cluster), we employ the $\Delta$-variance method to calculate the power spectra of the fractional fluctuations, $\delta y / y$ for SZ and $\delta S / S$ for X-ray. The $\Delta$-variance technique uses a Ricker wavelet filter to isolate fluctuations on a given spatial scale by computing the variance of the convolved image. The filter itself is constructed as the difference of two Gaussian kernels with slightly different widths, which naturally compensates for masked regions in the map \citep{Arevalo_PS}. For completeness, full details of the implementation are provided in Appendix \ref{sec:delta_variance_appendix}. 
\\\\
This procedure yields the 2D projected fluctuation power spectra, which we then deproject to obtain 3D power spectra. The mathematical details of the deprojection, following \citet{churazov_coma,Khatri_2016, romero2024}, are given in Appendix \ref{app:depro}. Throughout this work, we characterize fluctuations using the amplitude spectra, defined as
\begin{align}
A_{3D}(k) &= \sqrt{4\pi k^3 \mathcal{P}_{3D}(k)} \\
A_{2D}(k) &= \sqrt{2\pi k^2 \mathcal{P}_{2D}(k)} ,
\end{align}
where $k^{-1}$ is the spatial scale of fluctuations.
These amplitude spectra provide direct estimates of the amplitudes of pressure and density fluctuations within A3266.

\subsubsection{\sptplanck{} instrumental effects}
\label{sec:sz-ps}
To calculate the power spectra for the pressure fluctuations within A3266, we follow the steps outlined above. The power spectra are reconstructed over the range $2.8$\arcmin\ to $14.2$\arcmin\ (corresponding to $210$ kpc to $1000$  kpc), but non-zero fluctuations are only detected within the range of $8.6$\arcmin\ to $14.2$\arcmin\ (corresponding to $610$ kpc to $1000$  kpc). In addition, we apply a circular Hanning window to mitigate highly amplified map regions that would otherwise dominate the overall power spectrum. 
These regions are due to dividing by the beta-model, which attains very small values near the edges of the circular cutout. We compute the angular transfer function of the Hanning window for power spectra computed from the $\Delta$-variance method, and correct all derived power spectra for this transfer function. 
\\\\
To estimate the portion of the power spectrum attributable to both instrumental and astrophysical noise ($\mathcal{P}_\mathrm{3D, noise}$), and thus unrelated to the ICM, we apply the same method to residual fluctuations obtained from the random noise cutouts described in Sec.~\ref{sec:sz_noise}. Specifically, for each noise cutout we subtract the mean, normalize by the convolved beta-model fitted to the cluster, and then compute its $\Delta$-variance power spectrum. We repeat this process for the 100 randomly sampled noise cutouts. The mean power spectrum recovered from these 100 cutouts is then subtracted from the one obtained from the cutout centered on A3266 to yield the power spectrum due solely to ICM fluctuations ($\mathcal{P}_{\text{3D}}$). These noise spectra are also used to estimate uncertainties on $\mathcal{P}$, since they intrinsically contain all possible non-idealities such as pixel-pixel correlations. Finally, as described in Appendix~\ref{sec:xfer_function}, we correct for the effective transfer function of both the circular window and the SPT PSF, both of which have been empirically determined for the $\Delta$-variance method.

\subsubsection{\erosita{} instrumental effects}
\begin{figure}
    \centering
    \includegraphics[width=\linewidth]{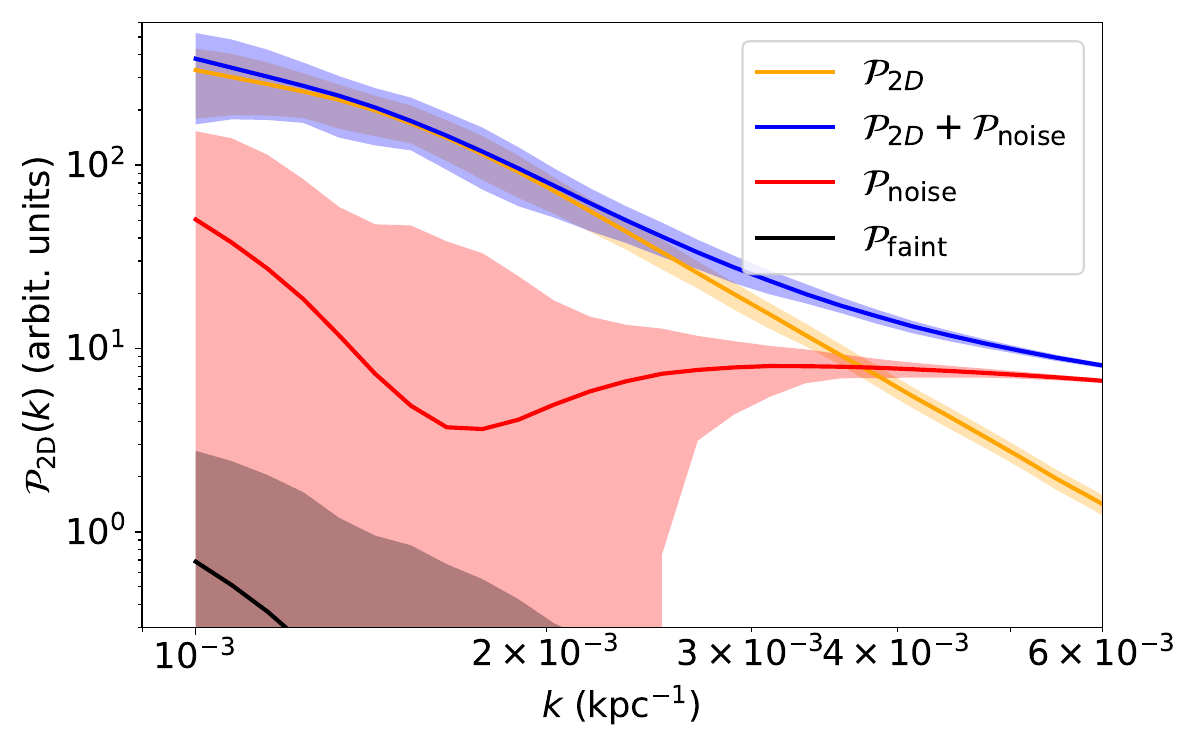}
    \caption{2D X-ray power spectra ($P_{2D}$), amplitude from noise images ($P_{\mathrm{2D,\ noise}}$), and amplitude from faint point sources ($P_{\mathrm{2D,\ faint}}$) as described in the text.}
    \label{fig:X_ray_noise}
\end{figure}

In addition to the steps described in Section \ref{PS}, we account for statistical and systematic uncertainties in the density fluctuations measurement \citep[e.g.,][]{churazov_coma,Zhuravleva_2014,romero_2023,Heinrich_mergers}.
We estimate the Poisson uncertainty at each wavenumber by generating 60 noise images, where the number of counts in each pixel is randomly drawn from a Poisson distribution based on the original X-ray image. 
The power spectrum of each noise image ($\mathcal{P}_\mathrm{2D,\ noise}$) is measured and the median is subtracted from the measured power spectrum, as done in subsection \ref{sec:sz-ps}. 
\\\\
To constrain the contribution to $\mathcal{P}_\mathrm{2D}$ from faint point sources ($\mathcal{P}_\mathrm{2D,\ faint}$), we first measure the power spectrum of our image without masking point sources ($\mathcal{P}_\mathrm{2D} + \mathcal{P}_\mathrm{2D,\ bright}$). Then, we measure the flux distribution of point sources in the entire image $dN/dF$, and fit this distribution to a single power law. Assuming $dN/dF$ does not change shape for faint point sources, we can use this distribution to find $\mathcal{P}_\mathrm{2D,\ faint}$ given
\begin{align}
    \mathcal{P}_\mathrm{2D,\ bright}/\mathcal{P}_\mathrm{2D,\ faint}=\left. \int_{F_\mathrm{min}}^{F_\mathrm{max}}\frac{dN}{dF} F^2 dF \middle/ \int_{0}^{F_\mathrm{max}}\frac{dN}{dF} F^2 dF \right.
\end{align}
where $F_\mathrm{max}$ and $F_\mathrm{min}$ are fluxes of the brightest and dimmest point sources in the \erosita{} image. We find $\mathcal{P}_\mathrm{2D,\ faint}$ is only comparable to $\mathcal{P}_\mathrm{2D}$ when $k^{-1} \lesssim 50$ kpc, and so it can be safely ignored in our calculations, as shown in Fig. \ref{fig:X_ray_noise}. Finally, we determine the effect of the \erosita{} PSF on our measured fluctuations by creating a blank image and randomly placing 100 point sources within it. We then measure the power spectrum of this image, along with the power spectrum of the image after it has been convolved with the \erosita{} PSF. The effect of the PSF on $\delta \rho/\rho_k$ is proportional to the square root of the ratio between the convolved and unconvolved power spectra, which we find to be $\lesssim 25\%$ at scales $\gtrsim 2.3$\arcmin (160 kpc). We therefore limit our measurement of $\delta \rho/\rho_k$ to scales larger than 160 kpc, and correct for the PSF in our final spectra.

\section{Results and Inferences} \label{result}
\begin{figure}
    \centering
    \includegraphics[width=\linewidth]{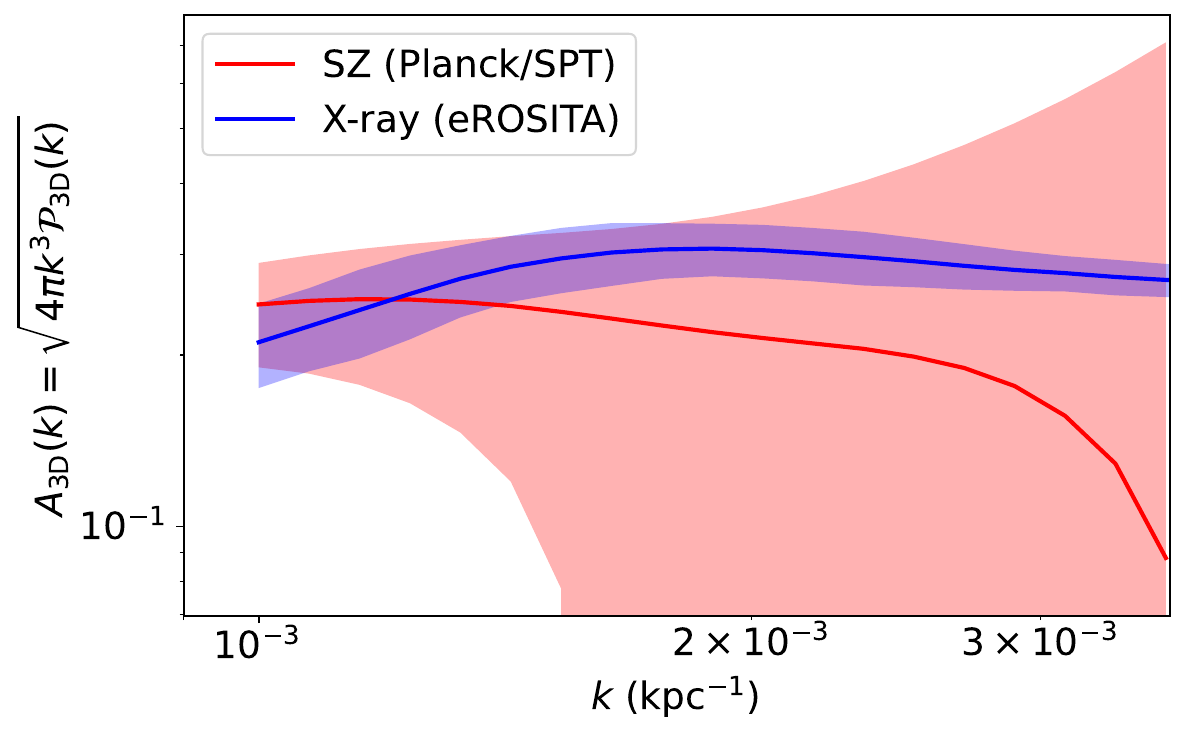}
    \caption{Measured SZ and X-ray 3D residual normalized fluctuation amplitude spectra for A3266 from our nominal analysis. Uncertainties are 1$\sigma$.}
    \label{fig:sz_and_xray_ellip_patch}
\end{figure}

With relatively large measurement uncertainties, particularly for scales smaller than $\approx 500$ kpc, the pressure fluctuation power spectrum is consistent with a constant value at all scales, see Fig. \ref{fig:sz_and_xray_ellip_patch}. Aggregating all scales, we find $A_{\text{3D}}(k) = 0.24 \pm 0.11$ for pressure. The density fluctuation power spectrum is also approximately constant with scale, although there is a broad peak for scales near $500$ kpc with a value of $0.31 \pm 0.03$.

\subsection{Effective equation of state of perturbations}
We compare the ratio of density and pressure fluctuation amplitudes, and note that they are consistent given the measurement uncertainties. Specifically, we find a weighted pressure/density amplitude ratio of $\zeta=1.00 \pm 0.55$, where we have excluded small-scale modes at $k^{-1} < 600$~kpc from this calculation since they are not well-constrained by the SZ data. The uncertainty on these values is dominated by the SZ measurement noise, with a negligible contribution from the X-ray errors. 
This value is consistent with the either isothermal or adiabatic perturbations, and inconsistent with isobaric perturbations at $\sim 2 \sigma$ significance.  It is important to note that this value includes all of the unmasked regions within a radius of 1 Mpc, and it is therefore likely that all three types of perturbation are present to some extent.

\subsection{Mach numbers and non-thermal pressure contribution}
In order to estimate the Mach number from the fluctuation amplitudes, they are measured at the same scale of $k^{-1}=500$ kpc motivated by the scale of the X-ray peak amplitude. We use the pressure fluctuation amplitude of $\delta P/P_k=0.22 \pm 0.27$ \footnote[1]{We note our formal uncertainty at this scale extends to unphysical regimes, which is then propagated to the Mach number constraints.} and the density fluctuation amplitude of $\delta\rho/\rho_k=0.31 \pm 0.03$ at this scale to obtain $\mathcal{M}_{\mathrm{1D},k}$ from the relation found by \citet{Irina_Mach}; 
\begin{align}\label{eqn:mach}
    \mathcal{M}_{\mathrm{1D},k} &= \delta\rho/\rho_k/\eta_\rho \\
     &= \delta P/P_k/\eta_P
\end{align}
where we use the scaling relation for clusters with unrelaxed morphologies ($\langle \eta_P\rangle = 1.5 \pm 0.5$ and $\langle \eta_\rho \rangle = 1.3 \pm 0.5$). To calculate the uncertainty on the Mach number, we include both the measurement uncertainties on the pressure and density amplitudes, along with the systematic uncertainties on the value of $\eta$ obtained from simulations. This gives us a $\mathcal{M}_{\text{1D},k} = 0.15 \pm 0.19$ and $\mathcal{M}_{\text{1D},k} = 0.24 \pm 0.09$ from pressure and density fluctuations, respectively. Both measurements indicate subsonic motions within the ICM. In addition, the uncertainties on the two Mach numbers are similar due to the large systematic uncertainty in the $\eta$ parameter obtained from simulations. We note that our measured values for the Mach number are similar to those obtained for clusters of comparable mass, redshift, and dynamical state range from the analysis of \cite{Dupourqu_2024}, and consistent with the median value obtained for the unrelaxed cluster sample studied by \cite{Heinrich_mergers}. 
\\\\
Assuming that the nonthermal pressure support in A3266 is primarily from turbulent motions, we can relate the Mach number to the nonthermal pressure ratio - 
\begin{align}
    \frac{P_{NT}}{P_{TOT}} = \bigg(\frac{\gamma \mathcal{M}_\mathrm{3D}^2}{3 + \gamma \mathcal{M}_\mathrm{3D}^2}\bigg)
\end{align}
where $\gamma=5/3$ is the adiabatic index.
From our measurements, we find non-thermal pressure support (the fraction of non-thermal pressure to total pressure) of $0.085 \pm 0.062$ from density fluctuations and $0.068 \pm 0.050$ from combining density and pressure fluctuations. Thus, we find that on average, $\approx 7\%$ of the total pressure support in the ICM for A3266 comes from non-thermal pressure support from turbulent motions.

\subsection{Fluctuation amplitude variations between hemispheres} 
\begin{figure*}
    \centering
    \begin{subfigure}[t]{0.49\linewidth}
        \centering
        \includegraphics[width=\linewidth]{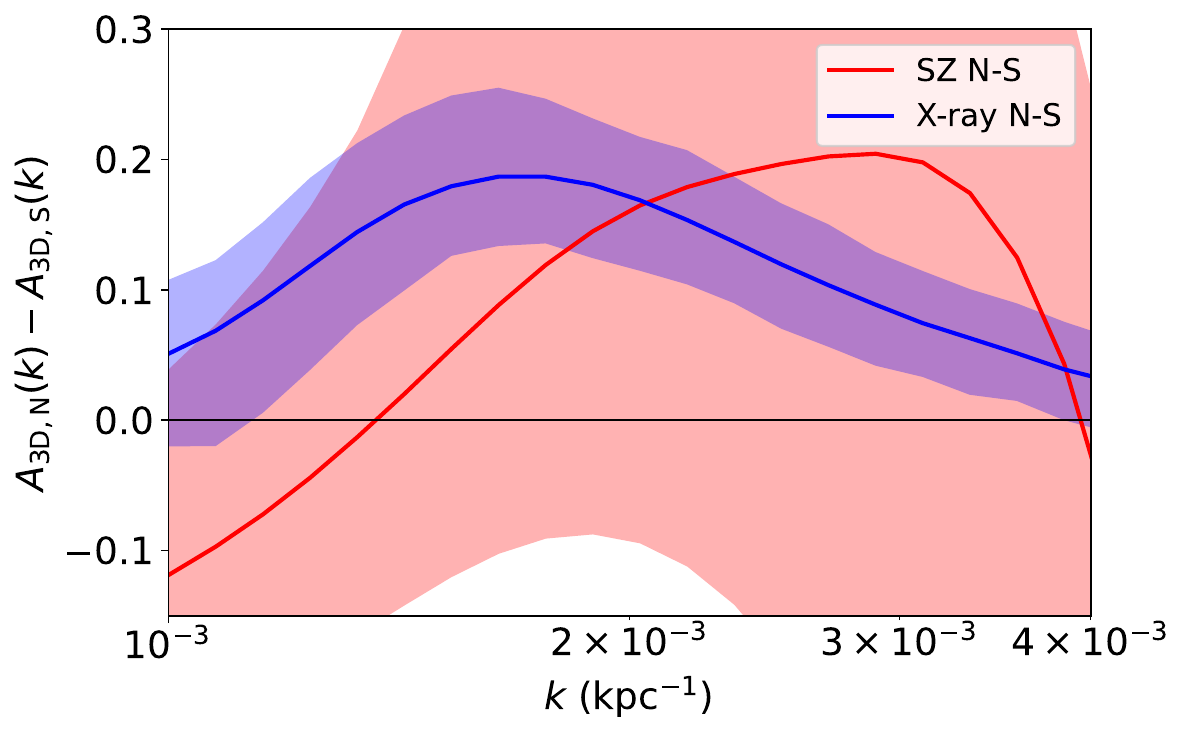}
        \caption{Difference between the 3D fluctuation amplitude for the northern and southern hemispheres for both the pressure and density fluctuations.}
        \label{fig:SZ_N_S_comp}
    \end{subfigure}
    \hfill
    \begin{subfigure}[t]{0.49\linewidth}
        \centering
        \includegraphics[width=\linewidth]{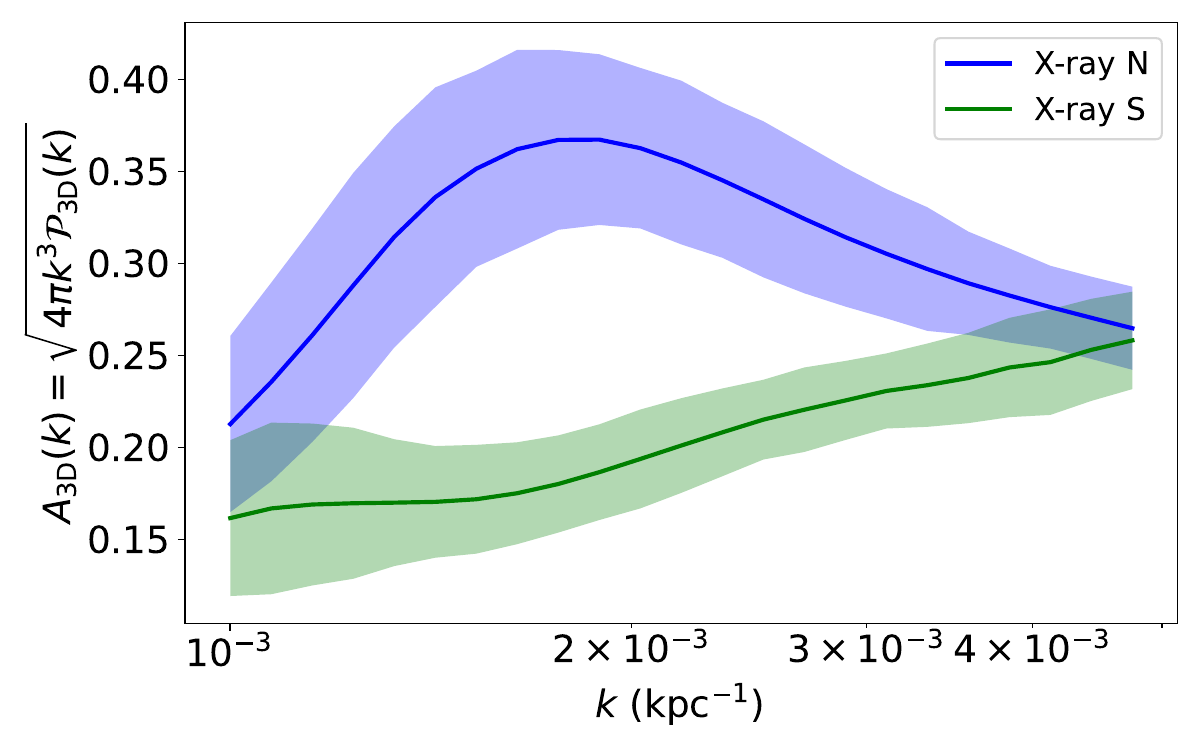}
        \caption{3D amplitudes of X-ray fluctuations from the North and South hemispheres of A3266.}
        \label{fig:Xray_N_S_comp}
    \end{subfigure}

    \caption{SZ and X-ray fluctuation amplitudes compared for the N and S hemispheres for the 9' patched model.}
    \label{fig:SPT_Xray_NS}
\end{figure*}

We perform a fluctuation analysis in different hemispheres of the cluster, splitting it into northern (N) and southern (S) halves, and compare the pressure and density amplitudes for each hemisphere. This hemisphere split is motivated by evidence of active merger processes happening in the NW and the NE of the cluster, as detailed in Section \ref{intro}. For the density fluctuation amplitudes from X-rays, we find consistent 3D fluctuation amplitudes at the largest and smallest scales probed, but at intermediate scales the northern hemisphere has a significantly higher amplitude, see Fig. \ref{fig:Xray_N_S_comp}. The 3D pressure amplitudes for each half are also shown in Fig. \ref{fig:SZ_N_S_comp}. We find no evidence for a statistically significant difference between them, given large measurement uncertainties. 
{We additionally search for differences in the density fluctuation amplitudes from X-rays between the east and west hemispheres, as well as the four quadrants of the cluster, finding no significant difference beyond the N-S variation.}

\subsection{Radial profile of density fluctuations}
\begin{figure}
    \centering
    \includegraphics[width=\linewidth]{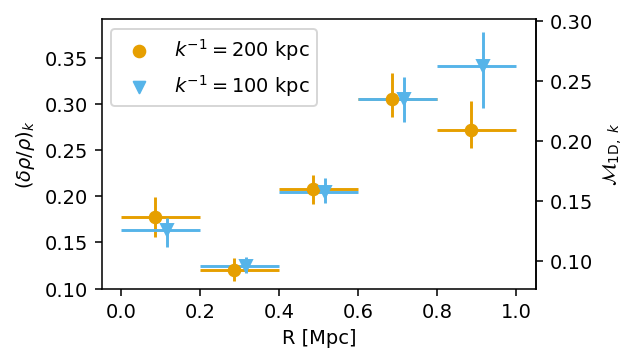}
    \caption{Radial profile of density fluctuations measured at spatial scales of 100 (blue triangles) and 200 kpc (orange circles).}
    \label{fig:radial_prof}
\end{figure}

To further investigate the distribution of fluctuations in A3266, we measure $\delta \rho/\rho_k$ in five 200 kpc-wide annuli from the cluster center out to 1 Mpc in radius. 
Due to the quality of the SZ data, a similar analysis of pressure fluctuations is not possible. 
Therefore, in this analysis, we do not remove the regions associated with the bright point source and scanning artifact in the \sptplanck\ map.
In Figure \ref{fig:radial_prof}, we present density fluctuations measured on scales of $k^{-1}=200$ and $100$ kpc as orange circles and blue triangles, respectively.
As each annulus is 200 kpc wide, the scales measured on 200 kpc may be suppressed due to insufficient sampling of these large-scale modes.
The 100 kpc modes are well sampled, however we note that the effect of the \erosita{} PSF is $\sim 35\%$ at this scale, which is corrected for.
Additionally, the outer three annuli have significant influence from faint point sources at this scale.
{Despite these possible systematics, the overall shape of the radial profile does not change significantly between $k^{-1}=200$ and $100$ kpc, suggesting any such systematic bias is small compared to the measurement uncertainties}. 
We additionally calculate the 1-dimensional Mach number ($M_{\mathrm{1D,}k}$) using Equation \ref{eqn:mach}.
\\\\
This profile shows an interesting, non-monotonic trend with radius. Fluctuations begin at a moderate value of $\sim 0.17$, before decreasing, then rising out to at least 0.8 Mpc. Between 0.8 and 1 Mpc, it is unclear whether fluctuations flatten out at $\delta\rho/\rho_k\approx 0.3$ or continue increasing.
This can be interpreted as evidence of an inhomogeneous velocity field out to at least a radius of $\sim 0.6$ Mpc, {as a single driver producing a homogeneous velocity field would result in a flat $\delta \rho/\rho_k$ profile}. 
The enhanced fluctuations/velocity in the innermost region can be explained by the stripping of gas from the infalling substructure. 
The rising fluctuations towards the cluster outskirts are expected, given the increased influence of accretion towards cluster outskirts.
This trend is also consistent with cosmological simulations \citep{Irina_Mach}.
We note, however, that the fluctuations measured in that work are scale-independent. 
While the comparison between these values and our scale-resolved measurements is not one-to-one, it is appropriate given that $\delta\rho/\rho_k$ is dominated by large scales.
The likely explanation for these increasing fluctuations is that the effects of mergers and accretion from filaments are more prominent at larger radii. A possible explanation of the merger-driven fluctuations lasting longer in cluster outskirts is that the eddy turnover time (proportional to the dynamical time) is longer in the outskirts, therefore it takes longer for the turbulence induced fluctuations to dissipate \citep{Gaspari_cascade}.
\\\\
{It is interesting to compare this profile to the one obtained from \textit{XRISM} observations of the relaxed cluster A2029 \citep{xrism_a2029}, where the velocity dispersion was found to decrease with radius out to $R_\mathrm{2500}$ ($\sim 670$ kpc.)}
Comparing these results is once again nontrivial due to the different nature of velocities measured via line broadening versus fluctuations.
While the measurements presented here are deprojected and multi-scale, velocity dispersions from \xrism{} are projected along the line-of-sight and associated with an effective length scale \citep{xrism_perseus} determined by the X-ray emissivity profile of the cluster.
Due to this emissivity weighting effect, the A2029 velocity dispersion measurements from \textit{XRISM} decrease in amplitude and increase in scale in each successive pointing further from the cluster center.
Assuming the velocity power spectrum follows a cascade that does not drastically change shape from the cluster center to $R_\mathrm{2500}$, the increasing scale indicates that the decrease in velocity dispersion observed with \textit{XRISM} is even more severe than it appears.
\\\\
In this work we trace the velocity fluctuation power spectrum amplitude at a single scale, initially seeing a decreasing trend with radius, consistent with A2029. 
Outside of 400 kpc, this trend reverses and velocity fluctuation power spectrum amplitude increase out to at least 800 kpc.
It is possible that XRISM would observe a similar reversal in A2029, where the velocity dispersion increases with radius, if measurements were made at larger radii.

\section{Discussion} \label{impact}
\subsection{Impact of model choice}
First, to assess the impact of patching, we compute a pressure to density fluctuation ratio of $1.19 \pm 0.55$ in the no patching case for an elliptical model. Comparing to our fiducial value, we note that the shift in the central value is lower than our uncertainty implying negligible impact on our results due to patching. 
We also assess the impact of utilizing a spherical, rather than an elliptical model. We also do not include a patch with this spherical model, to make it as different from our nominal case as possible. For the SZ data, the power spectrum does not change in a statistically significant manner, suggesting that any artificially-induced large-scale fluctuations due to intrinsic ellipticity are small compared to the measurement uncertainties. In contrast, the density fluctuation power spectrum obtained from the spherical model is higher than the one obtained from the elliptical model at modest statistical significance, particularly at large scales where it is boosted by approximately $1.6\sigma$. We find a weighted pressure/density amplitude ratio of $0.85\pm 0.43$ for a spherical model with no patching. Again, we note consistency with our fiducial value given uncertainties. Thus, we find that we are dominated by the uncertainty in SZ data over systematic uncertainties in our model choice. The pressure and density fluctuation amplitude plots for the unpatched models and spherical models are given in Appendix \ref{sec:model_impact_ps}.

\subsection{Potential improvements from new SZ instrumentation}
\begin{figure}
    \centering
    \includegraphics[width=0.97\linewidth]{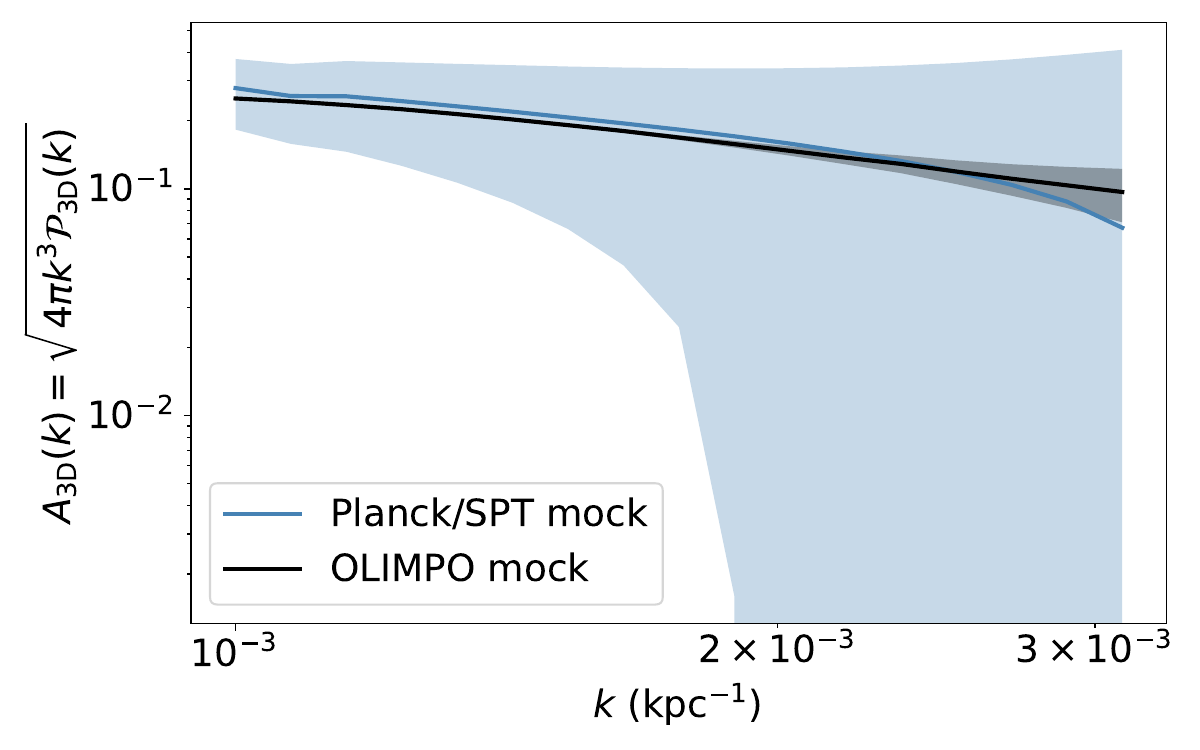}
    \caption{3D de-projected power spectra with error-bars from both SPT and OLIMPO like noise patches.}
    \label{fig:olimpo_mock}
\end{figure}

It is difficult to measure ICM fluctuations from existing state of the art SZ data due to a relative lack of sensitivity compared with X-ray observations. Furthermore, ground-based SZ observations, both from surveys and from targeted observations, suffer from atmospheric brightness fluctuations that are difficult to disentangle from ICM fluctuations, particularly at angular scales larger than a few arcminutes. While we utilize the combined \sptplanck{} map for this analysis, most of the sensitivity to scales larger than $\sim 350$ kpc (where pressure fluctuations are detected) is provided by \planck{} \citep{Bleem_2022}. While the SPT data have lower noise than the \planck\ data at small angular scales, the typical pressure fluctuation power spectrum falls rapidly with increasing $k$, thus making it undetectable with the existing SPT data.
\\\\
To provide context for what would be possible from upcoming SZ instrumentation, we consider OLIMPO, which is a proposed Antarctic balloon-borne SZ imager capable of delivering far deeper images than those utilized in this analysis \citep{Sayers_2024_OLIMPO}. To determine how well OLIMPO could probe pressure fluctuations, and to place them in context with existing data, we generate mock observations of A3266 with Kolmogorov fluctuations with a fractional amplitude of 0.2 from both OLIMPO and \sptplanck\ from noise estimates for each instrument. In Fig. \ref{fig:olimpo_mock} we plot the 3D de-projected power spectra, which indicate that OLIMPO provides far better sensitivity, with a factor of $\approx 130$ lower noise at the largest scales and a factor of $\approx 13$ lower noise at the smallest scales. Combined SZ/X-ray analyses from such data will provide far more incisive tests of ICM turbulence, including a definitive assessment of the effective EOS of perturbations and the ability to probe a wide range of physical scales in both observables.

\section{Conclusion}
{We perform a density and pressure fluctuation measurement on A3266, a low-redshift, massive cluster using state of the art \sptplanck\ and \erosita\ data.  Our analysis covers the area from the cluster center out to a radius of 1 Mpc.}
\begin{itemize}
    \item 
    {We detect pressure fluctuations on spatial scales between $\sim 600$ and $ 1000$ kpc, with measurements on smaller scales limited by instrumental noise. 
    Density fluctuations are measurable from $\sim 160$ to $\sim 1000$ kpc, where the limiting factor is the \erosita{} PSF.
    The significant difference in the scales probed and the uncertainty in the measurements, which are shown in Fig. 2, highlights the need for deep, high-resolution SZ imaging.}

    \item 
    {By calculating the ratio of pressure to density fluctuations, we measure the effective EOS parameter of fluctuations in the ICM to be $\zeta=1.00\pm 0.55$, consistent with values for isothermal and adiabatic fluctuations. The uncertainty in this parameter is driven by the SZ observations, despite our measurement of pressure fluctuations being comparable to the most sensitive measurements in the literature \citep{Khatri_2016,romero_2023,romero2024,romero2025szxraysurfacebrightnessfluctuations}. 
    }

    \item
    {We measure fluctuations in the northern and southern hemispheres of A3266, finding significantly higher $\delta\rho/\rho_k$ in the north compared to the south. 
    We do not detect a difference in $\delta P/P$ between the hemispheres.
    The enhanced density fluctuations in the north are likely due to ongoing and previous merger activity, as the ongoing NE-SW merger and the previous SE-NW merger have recently disrupted the northern half of the cluster.}

    \item 
    We produce a radial profile of density fluctuations by measuring the density fluctuations power spectrum in five 200 kpc-wide annuli.
    This profile reveals a non-monotonic trend, wherein $\delta\rho/\rho_k$ initially drops outside of 200 kpc, then increases in amplitude out to at least 800 kpc.
    As $\delta\rho/\rho_k$ traces the Mach number of gas motions in the ICM, this varying profile indicates an inhomogeneous velocity field in A3266.
    The enhanced fluctuations within a radius of 200 kpc from the cluster center may be due to the ram-pressure-stripping of the infalling subcluster, while the increasing Mach number (and non-thermal pressure) towards the cluster outskirts may indicate the increasing effects of accretion or lingering effects of previous mergers. Additionally, the initial drop is qualitatively consistent with \xrism{} measurements of velocities in A2029, while the increase towards larger radii is consistent with predictions from cosmological simulations.

    \item 
    Using scaling relations between density/pressure fluctuations and gas velocities, we obtain the turbulent Mach number measured by the density and pressure fluctuation amplitude at the X-ray peak to be $\mathcal{M}_{\text{1D,SZ}} = 0.15 \pm 0.19$ and $\mathcal{M}_{\text{1D,X}} = 0.24 \pm 0.09$. These are consistent with other fluctuation studies for clusters within a similar mass, redshift and dynamical state range. Assuming an EOS parameter of $5/3$ for the gas, we derive the non-thermal pressure support to be $0.085 \pm 0.062$  from density fluctuations and $0.068 \pm 0.050$ from combined pressure and density fluctuations. 
    
\end{itemize}

A range of SZ facilities that are now, or soon will be, online, will provide an order of magnitude improvement in sensitivity compared to existing data \citep{Benson_2014, SO, Wilson2020}. However, all of these facilities operate from the ground, and will thus lack sensitivity to the largest scales within the ICM, particularly for lower redshift clusters. Possible future instruments like OLIMPO, which would operate above the atmosphere, would be ideally suited to fluctuation measurements and capable of measuring them to at least 2 Mpc in radius \citep{Sayers_OLIMPO_mmuniv}. On the X-ray side, planned upcoming missions such as \textit{NewAthena} offer the promise of large collecting area, small PSF width, and large FOV, and would provide deep enough observations to measure SB fluctuations extending to at least 2-3 Mpc in radius \citep{newATHENA}.

\begin{acknowledgments}
H. Saxena, J. Sayers, and D. White were partially supported by NASA award 80NSSC25K0597. A. Heinrich and I. Zhuravleva were partially supported by NASA award 80NSSC24K1488. E. Bulbul acknowledge financial support from the European Research Council (ERC) Consolidator Grant under the European Union’s Horizon 2020 research and innovation program (grant agreement CoG DarkQuest No 101002585). The authors would like to acknowledge Charles Romero for insightful discussions regarding the $\Delta$-variance method and associated systematics. 

\end{acknowledgments}

%



\appendix

\section{$\Delta$-variance method}
\label{sec:delta_variance_appendix}
Here, we present the details of the $\Delta$-variance method as compiled from \citet{Arevalo_PS} and \cite{Khatri_2016}. The method relies on computing the power spectrum for a broad interval of wavenumbers, allowing for (and correcting for) gaps in the data. We consider a image defined as $I(x)$, where $x$ is a two-dimensional pixel location, and a corresponding boolean mask $M(x)$, which is 0 for the regions that are masked and 1 everywhere else. We define a normalized gaussian filter in the spatial domain 
\begin{align}
    G_{\sigma} (x) = \frac{e^{-x^2/2\sigma^2}}{2\pi \sigma^2}.
\end{align}
The filter is then defined as the combination of two gaussians that isolates fluctuations at a given scale
\begin{align}
    F_{k} (x) &= G_{\sigma_1}(x) - G_{\sigma_2}(x) \\
    \sigma_1 &=\frac{0.225}{k \sqrt{1+\epsilon}} \\
    \sigma_2 &=\frac{0.225}{k}
\end{align}
as derived in \citet{Arevalo_PS}, with $\epsilon = 0.001$. In the case of gaps in the data, we convolve the image and the mask with the two gaussian filters above, subtract the results, and apply the original mask to obtain 
\begin{align}
    S_{k}(x) = \bigg( \frac{G_{\sigma_1} * I}{G_{\sigma_1} * M} - \frac{G_{\sigma_2} * I}{G_{\sigma_2} * M}\bigg)M(x).
\end{align}
The square of this is then integrated, and the variance is computed by rescaling with the number of pixels that are masked 
\begin{align}
    V_{k} = \frac{N}{N_{M=1}}\int S^2_{k} (x) d^2x.
\end{align}
Finally, the 2D power spectrum $\mathcal{P}_{k}(x)$ is obtained by dividing the variance by the integrated filter normalization. We deviate from the analytical approximation of the normalization of the variance as presented in \cite{Arevalo_PS} and use the numerical integral of the filter on a discrete 2D grid as the normalization of the variance to obtain the power spectrum. 

\section{Deprojection to 3D}
\label{app:depro}
We follow \citet{churazov_coma}  in their methodology to deproject the power spectra from 2D to 3D, first focusing on the SZ data as a specific example. In this formalism, the 2D and 3D power spectra are related through the window function 
\begin{align}
    \mathcal{P}_{2D}(k) &= \int \mathcal{P}_{3D}(k, k_z)|\tilde{W}(k_z)|^2 dk_z \\
\end{align}
where the window function for pressure fluctuations is 
\begin{align}
    W(r_e, z) &= \frac{\sigma_T}{m_e c^2}\frac{\bar{P}(r_e,z)}{\bar{y}(r_e)}
\end{align}
where $z$ is along the line of sight, $r_e$ is the elliptical radius on the plane of the sky, $\bar{y}$ is the smooth $\beta$-model, and $\bar{P}$ is the 3D unperturbed model defined as 
\begin{align}
    \bar{P}(r_e) &= {P_0}{(1 + r_e^2+z^2)^{-3\beta}} \\
    P_0 &= \frac{m_e c^2 S_0}{\sigma_T r_c}\frac{\Gamma(3\beta)}{\sqrt{\pi} \Gamma(3\beta - 0.5)}
\end{align}
where $\Gamma$ is the gamma function, and the other parameters are as defined for the $\beta$-model above. 
Above some cutoff wavenumber, the window function begins to drastically drop off, and we thus approximate the 3D power spectrum as 
\begin{align}
    \mathcal{P}_{2D}(k) &\approx \mathcal{P}_{3D}(k) N(r_e)  \\
    N(r_e) &= \int |\tilde{W}(k_z)|^2 dk_z \\
    N_{\text{eff}} &= \langle N(r_e) \rangle \ 
\end{align}
where the average in the last line is taken until the radius corresponding to our window, which is equal to 1000 kpc. This $N_{\text{eff}}$ is then used to obtain the 3D power spectrum from the 2D power spectrum for the pressure fluctuations. 

To deproject density fluctuations, the window function is a three-dimensional $\beta$-model. In this calculation, we preserve the radial variation of the deprojection factor, which becomes 
\begin{align}
   N(r_e)= 2^{1-6\beta}\ \frac{\Gamma(3\beta)\ \Gamma(6\beta-\half)}{\Gamma(3\beta-\half)^2\ \Gamma(3\beta+\half)}\left(r_c\sqrt{1+({r_e}/{r_c})^2}\right)^{-1}
\end{align}
This deprojection factor is included in the power spectral calculation \citep{Heinrich_mergers}. \cite{Romero2024_Forecast} show that they find no statistically significant differences between these two methods using simulations.

\section{Corrections to the SZ power spectrum amplitude}
\label{sec:xfer_function}
\begin{figure}
    \centering
    \includegraphics[width=0.5\linewidth]{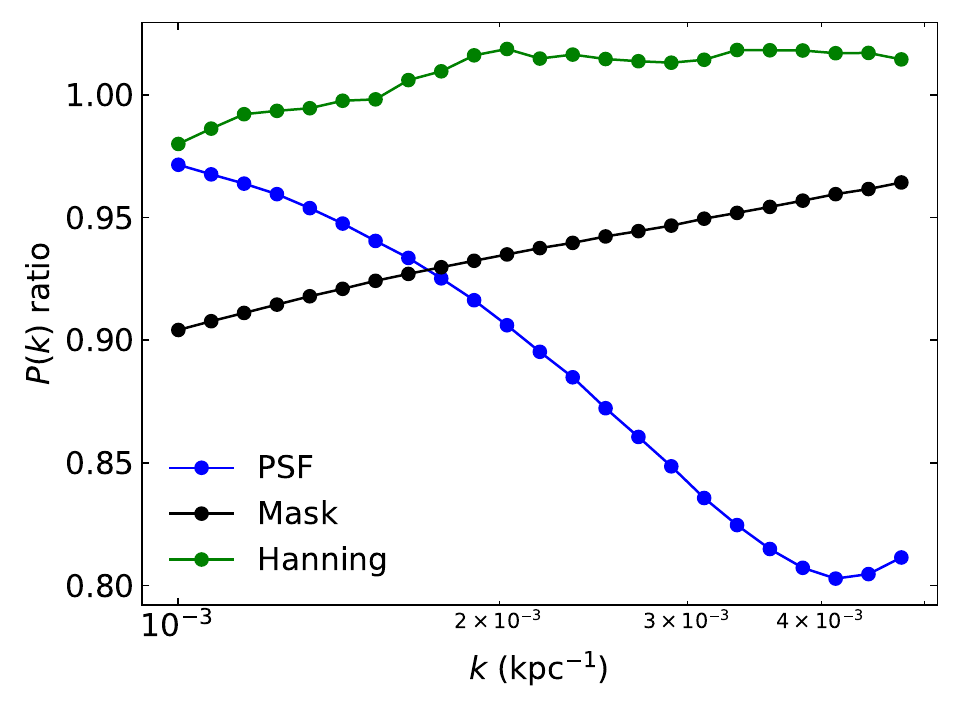}
    \caption{SZ power spectrum corrections applied in the pipeline due to PSF corrections, masking and smoothing corrections, and Hanning window effects, see the text for details.}
    \label{fig:PS_corrections}
\end{figure}

To determine any scale-dependent bias in the recovered pressure fluctuation amplitude, we perform an analysis of mock observations of a cluster with a fixed profile shape. This shape is assumed to be an elliptical $\beta$-model, similar to the best-fit model of A3266, to which we add Gaussian fluctuations and then convolve the result with the PSF shape of the \sptplanck\ map. To reduce the impact of noise on amplitude recovery, we select a random noise cutout from the set described in Sec.~\ref{sec:sz_noise} and multiply it by $10^{-6}$ to significantly suppress the noise amplitude prior to adding the cluster model to the noise cutout. We then analyze the resulting mock image with our standard analysis pipeline, which includes scale-dependent corrections for multiple steps, as illustrated in Fig. \ref{fig:PS_corrections}. First, the correction for signal attenuation at small scales (large $k$) due to the PSF differs from the idealized case, since the power spectrum estimated from the $\Delta$-variance method is not identical to that obtained from an FFT. The ratio of these two power spectra is shown in blue, and is applied as an additional correction to the transfer function of the idealized PSF. Next, we assess the effect of circular masking on power spectrum recovery. For this, we generate a pure Gaussian fluctuation field, apply the circular mask and \sptplanck\ PSF, and compare the input and output power spectra obtained from the $\Delta$-variance method. This yields a small correction, denoted as the Mask correction in Fig. \ref{fig:PS_corrections}. Finally,  we assess whether an additional correction factor is required due to utilizing a Hanning window in combination with the $\Delta$-variance method. Averaging over 1000 mock observations, we find deviations of less than $\simeq 0.02$ symmetrically distributed about a value of 1. Given that these deviations are negligible compared to our measurement uncertainties, we do not correct for them in our final analysis.

\section{Impact of model choice on power spectra}
\label{sec:model_impact_ps}

\begin{figure*}
    \centering
    \begin{subfigure}[t]{0.45\linewidth}
        \centering
        \includegraphics[width=0.96\linewidth]{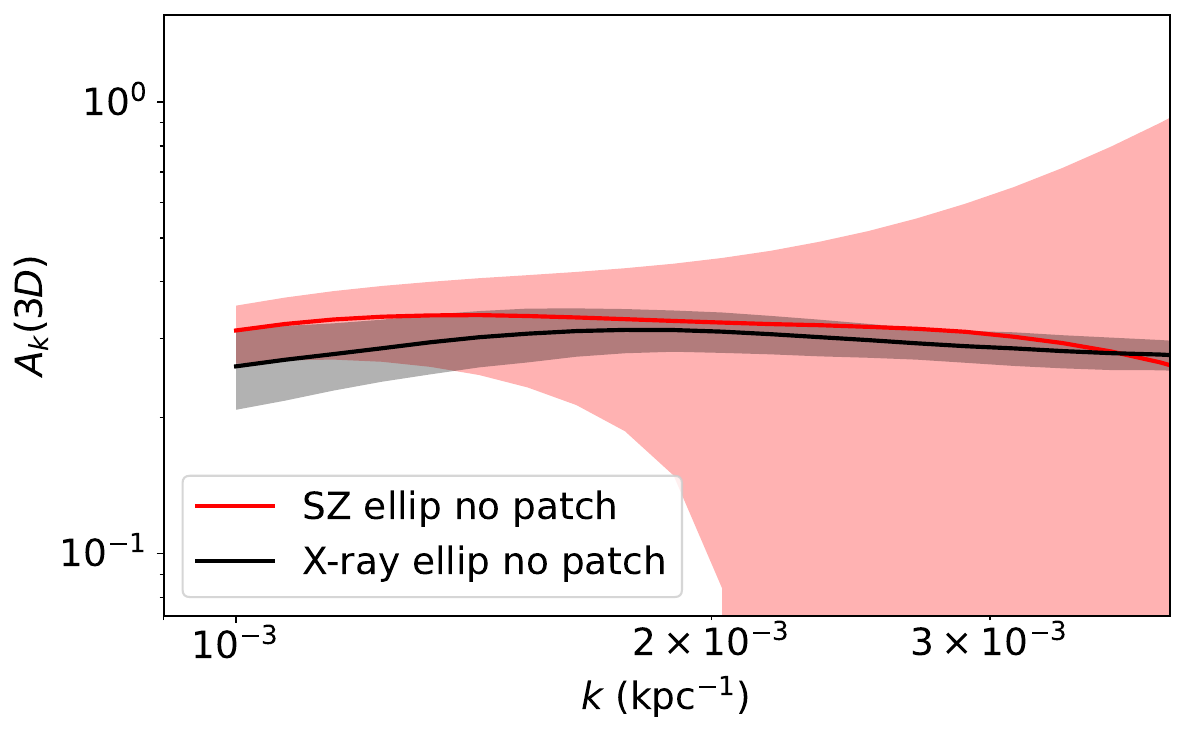}
        \caption{Pressure and density fluctuation amplitude comparison for elliptical model with no patching.}
        \label{fig:sz_and_xray_ellip_nopatch}
    \end{subfigure}
    \hfill
    \begin{subfigure}[t]{0.45\linewidth}
        \centering
        \includegraphics[width=\linewidth]{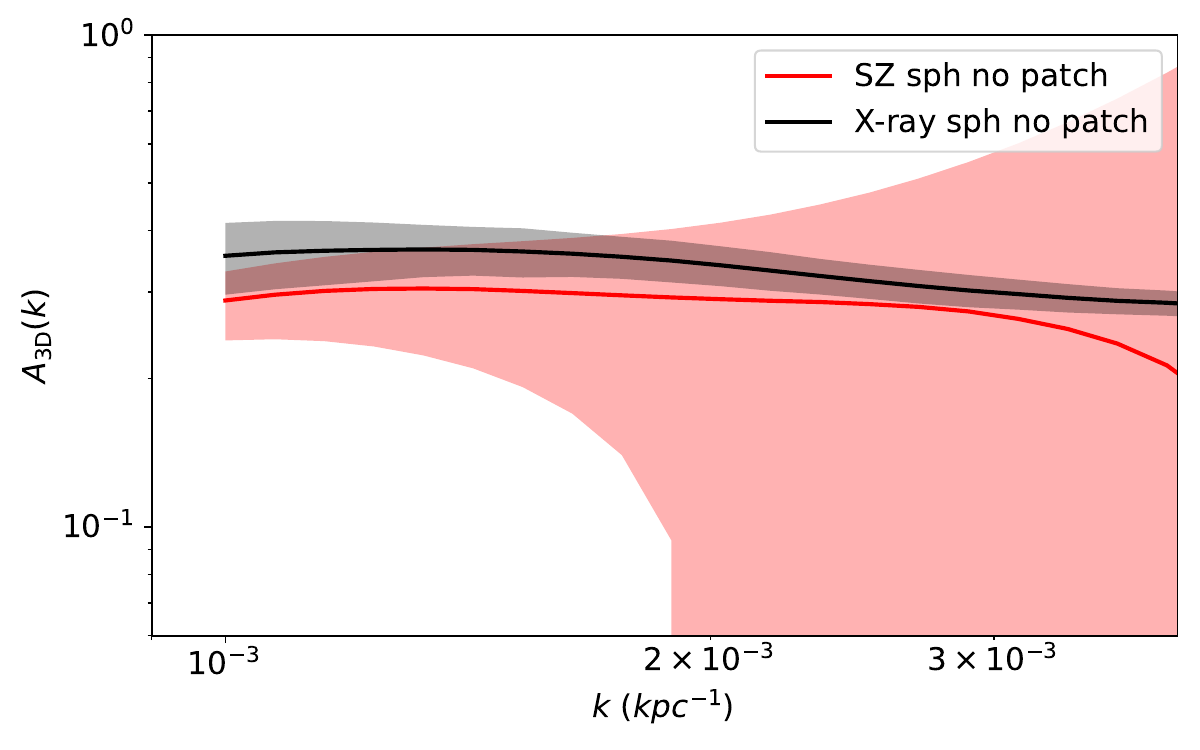}
        \caption{Pressure and density fluctuation amplitude comparison for a spherical model with 9\arcmin\ patching.}
        \label{fig:sz_and_xray_sph_patch}
    \end{subfigure}
\end{figure*}

The fluctuation power spectrum results obtained from an elliptical unpatched model are shown in Fig. \ref{fig:sz_and_xray_ellip_nopatch}. We note that the pressure fluctuation amplitude is higher than in the patched case (albeit within measurement uncertainty), suggesting that the impact of model choice is not important for the SZ data given the measurement uncertainties. In Fig \ref{fig:sz_and_xray_sph_patch}, we show the power spectra obtained from a spherical unpatched model, and note the increase at large scales of the X-ray power spectra as discussed in Section \ref{impact}. We also note that the peak of the X-ray power spectra shifts compared to those obtained from the nominal elliptical patched model. Specifically, it shifts from near 500 kpc in the nominal case to approximately 750 kpc in the spherical unpatched case.

\bibliography{sample7}{}
\bibliographystyle{aasjournalv7}



\end{document}